\documentclass[twocolumn,secnumarabic,amssymb, nobibnotes, aps, prd]{revtex4-1}
\usepackage[utf8]{inputenc}
\usepackage{epsf}
\usepackage{graphicx}
\usepackage{caption}
\usepackage{subcaption}
\usepackage{color}
\usepackage{soul}
\usepackage{gensymb}
\usepackage{float}
\usepackage{comment}

\begin{document}

\title{Effect of color reconnection and rope formation  on resonance production in p$-$p collisions in Pythia 8  }%


\author{Ankita  Goswami}
\email{ankita@phy.iitb.ac.in}
\affiliation{Indian Institute of Technology Bombay, Mumbai, 
  India-400076}

\author{ Ranjit Nayak }
\email{ranjit@phy.iitb.ac.in}
\affiliation{Indian Institute of Technology Bombay, Mumbai, 
  India-400076}

\author{Basanta Kumar Nandi}
\email{basanta@phy.iitb.ac.in}
\affiliation{Indian Institute of Technology Bombay, Mumbai, 
  India-400076}

\author{ Sadhana~Dash }
\email{sadhana@phy.iitb.ac.in}
\affiliation{Indian Institute of Technology Bombay, Mumbai, 
  India-400076}

\begin{abstract}
The resonance production in proton$-$proton collisions at $\sqrt{s}$ = 7 TeV and 13 TeV have been investigated using Pythia 8 event generator within the framework of microscopic processes like color reconnection and rope hadronization. Specifically, the observable effects 
of different modes of color reconnections on the ratio of yields of mesonic and baryonic resonances with respect to their stable counterpart 
have been explored as a function of event activity. A suppression in the ratio is observed as a function of number of multi-partonic interactions for mesonic resonances. The $\mathrm{\phi/K}$ and $\mathrm{\phi/\pi}$ ratios show an enhancement for high multiplicity events due to enhanced production of strange quarks via the microscopic process of rope hadronization in the partonic phase. The mechanism of the hadronization of color ropes together with the QCD-based color reconnection of partons predicted an enhancement in the ratio for baryonic resonances to non-resonance baryons having similar quark content. 
The yield ratios of resonances are found to be independent of the collision energy and strongly dependent on event activity.  
 \end{abstract}
\date{\today}
\maketitle

\section{Introduction}
Proton-proton (p$-$p) collisions have been studied extensively as a reference for the study of medium-induced phenomena in 
heavy ion collisions. However, recent measurements performed in high multiplicity p$-$p and proton-lead (p$-$Pb) collisions at LHC exhibited 
some features like strangeness enhancement \cite{nature}, long-range ridge-like structure in near-side of two particle correlations \cite{cms}, and collectivity effects \cite{flowcr} which are similar to that 
observed in heavy ion collisions. These interesting observations have  triggered the investigation of collectivity-driven features in small systems (like p$-$p) and is currently actively pursued. In high energy 
experiments at RHIC and LHC, the bulk properties of the medium formed due to collision of heavy ions have been widely studied to comprehend the 
conditions of late hadronic phase together with the effects of collective expansion from
initial stages. The hadronic resonances whose lifetime are shorter or comparable to the time scale for evolution of the fireball are
extremely sensitive to probe the properties of the dense medium \cite{reso1}. This might lead to
modifications in their production rate and characteristic properties (such as in vacuo mass and width)
due to dynamical interactions with the surrounding matter. The fraction of resonances that decays before the kinetic
freeze out (vanishing elastic interactions) are less likely to be reconstructed due to elastic scattering of
the decay products with the surrounding medium known as the resonance re-scattering effect. 
Alternatively, the  yield might increase in the hadronic phase via interactions known as resonance regeneration and can partially or 
fully compensate the yield loss due to re-scattering effect. The observation of suppression of the relative yield ratios of $\mathrm{K^{*0}/K}$ \cite{alice,auau} and $\mathrm{\Lambda(1520)/\Lambda}$ \cite{lambda} with centrality was attributed to the interplay of the medium-induced effects in the hadronic phase. The EPOS model estimations on hadronic resonances also described the suppression of $\mathrm{K^{*0}/K}$ in central Pb$-$Pb collisions due to  interactions in the extended hadronic phase \cite{epos1,epos2}. However, recent observation of  suppression of $\mathrm{K^{*0}/K}$ in high multiplicity p$-$p collisions \cite{anders} has triggered discussions regarding the role of other microscopic mechanisms present in the partonic phase. The p$-$p system due to its  smallness is not expected to have a medium or active hadronic phase to account for the observed suppression. The study of such small systems provide insights about the degree of similarity of the hadronization process to the  one present in larger systems. They can shed light on the contributions from initial partonic phase. Recently, the microscopic processes like color reconnections and rope hadronization were able to describe the  observation of  
strangeness enhancement \cite{strange,strange1}, near side long-range ridge-like correlations \cite{shoving}, and flow-like effects \cite{flowcr} in p$-$p collisions at LHC energies. In a recent study done in p$-$p collisions using the Pythia 8 generator \cite{pythia8}, the QCD-based color reconnection mechanism was able to reproduce the suppression trend in $\mathrm{K^{*0}/K}$ ratio as a function of event activity \cite{resoprd}. 
Color reconnections refer to the mechanism of formation of hadronizing strings between the outgoing partons regardless of their formation history. The Pythia 8 model implements the mechanism on the basis of minimization of the total string potential energy stored in these strings. The final outgoing partons generally tend to get color correlated with 
the partons which are nearer to them in momentum space resulting in minimizing the string length. Consequently, the total potential energy stored in these strings is also minimized. The details of the CR mechanism can be found in Ref. \cite{colorreco1,colorreco2}. In Pythia 8, the QCD-based color reconnection(CR((1)) allows the formation of junctions where three color lines join and can result in an enhancement of baryons. 
For high multiplicity p$-$p collisions, it may happen that several strings are very close to each other and can mutually interact among themselves 
 to form color ropes. These ropes have a larger effective string tension  and are likely to fragment into more strange quarks and diquarks which 
 leads to an enhanced  production of baryons and  strange hadrons \cite{rope1,rope2}.
The suppression of mesonic resonances using Pythia 8 was attributed to the mechanism of QCD-based color reconnection \cite{resoprd}. 
The color reconnections lead to the formation of shorter string lengths  which are more likely to fragment into less massive particles.  As resonances are massive, there is a decrease in the resonance to non-resonance ratio for high multiplicity events where the effect of CR 
is more noticeable. 
In this work, the study of resonance to non-resonance ratios has been carried out to investigate the effects of color reconnections and rope hadronization on mesonic and baryonic resonances using Pythia 8.  

\section{Pythia 8 Simulations}
The study is based on 20 million events of inelastic p$-$p collisions generated at $\sqrt{s}$ = 7 TeV and 13 TeV each using Pythia 8 event generator. The Monash 2013 tune \cite{monash} was used to simulate the events with multi-partonic interactions (MPI),  rope hadronization (RH) and two different modes of color reconnections (CR(0) and CR(1)). The production of various mesonic resonances like $\mathrm{K^{*}}$, $\mathrm{\phi}$, $\mathrm{\eta^{'}}$, and $\mathrm{\rho}$ as well as the baryonic resonances like $\mathrm{\Delta^{+}}$, $\mathrm{\Sigma^{*0}}$, and  $\mathrm{\Xi^{*0}}$, specifically the ratios of their yield to the yield of their stable counterpart (having same quark content) is studied as a function of event activity. The ratios of the considered resonance particles with their stable counterpart were obtained for $|\eta| < 0.5$ for different configurations. The configurations involved various combinations of the different modes of CR with (and without) the formation of ropes. As the observed charged particle multiplicity is strongly correlated with the number of MPIs ($N_{MPI}$), the effect of various mechanisms on resonance to non-resonance ratio is studied as a function of number of MPI's (or event activity). The observed variation with respect to $N_{MPI}$ is expected to yield similar trend as a function of charged particle multiplicity and is advantageous as the predictions can be directly compared with experimental results.

\begin{figure*}
\centering
\begin{subfigure}[b]{0.32\textwidth}
\includegraphics[width=\textwidth]{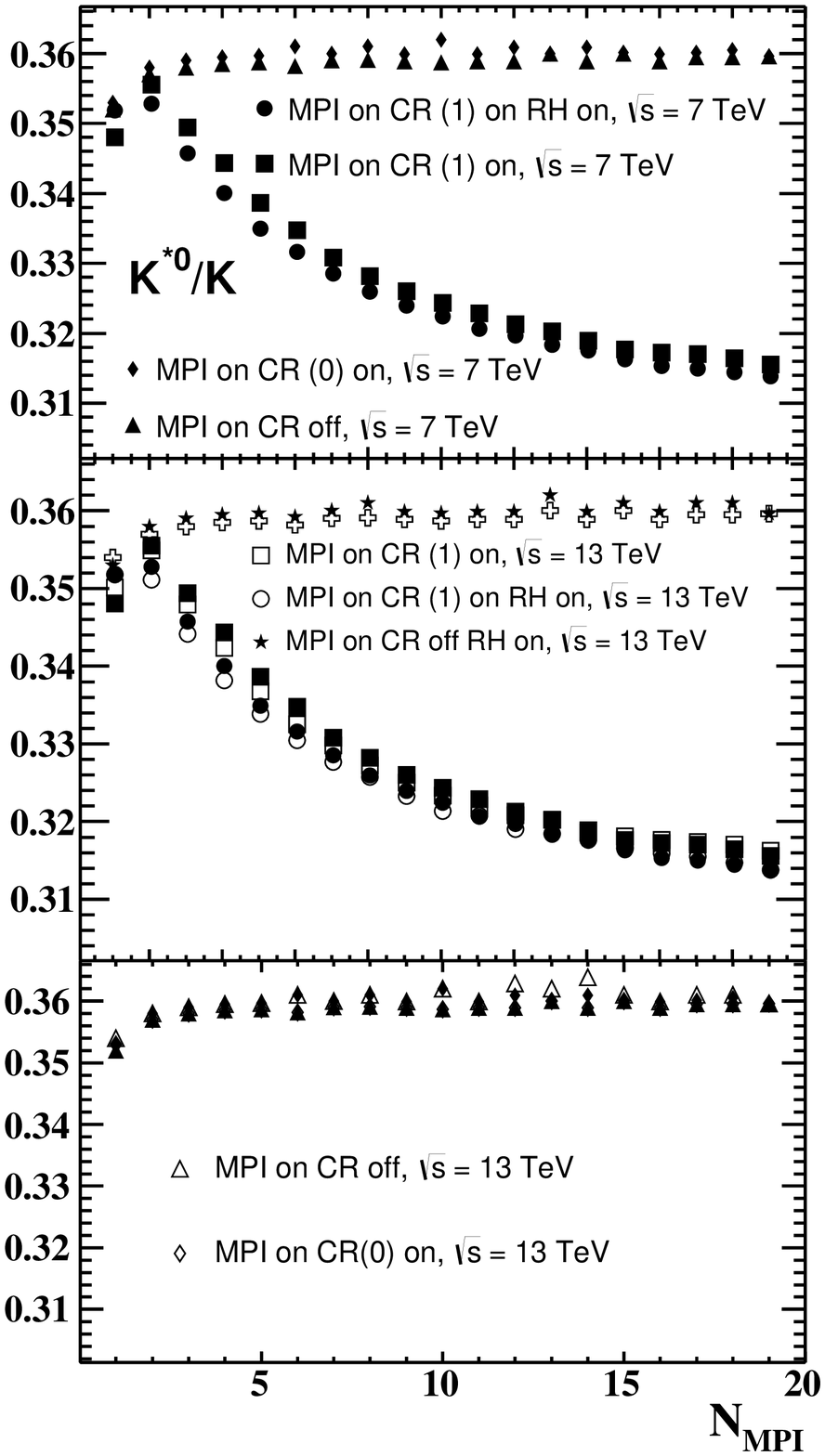}
\caption{}
\end{subfigure}
\begin{subfigure}[b]{0.32\textwidth}
\includegraphics[width=\textwidth]{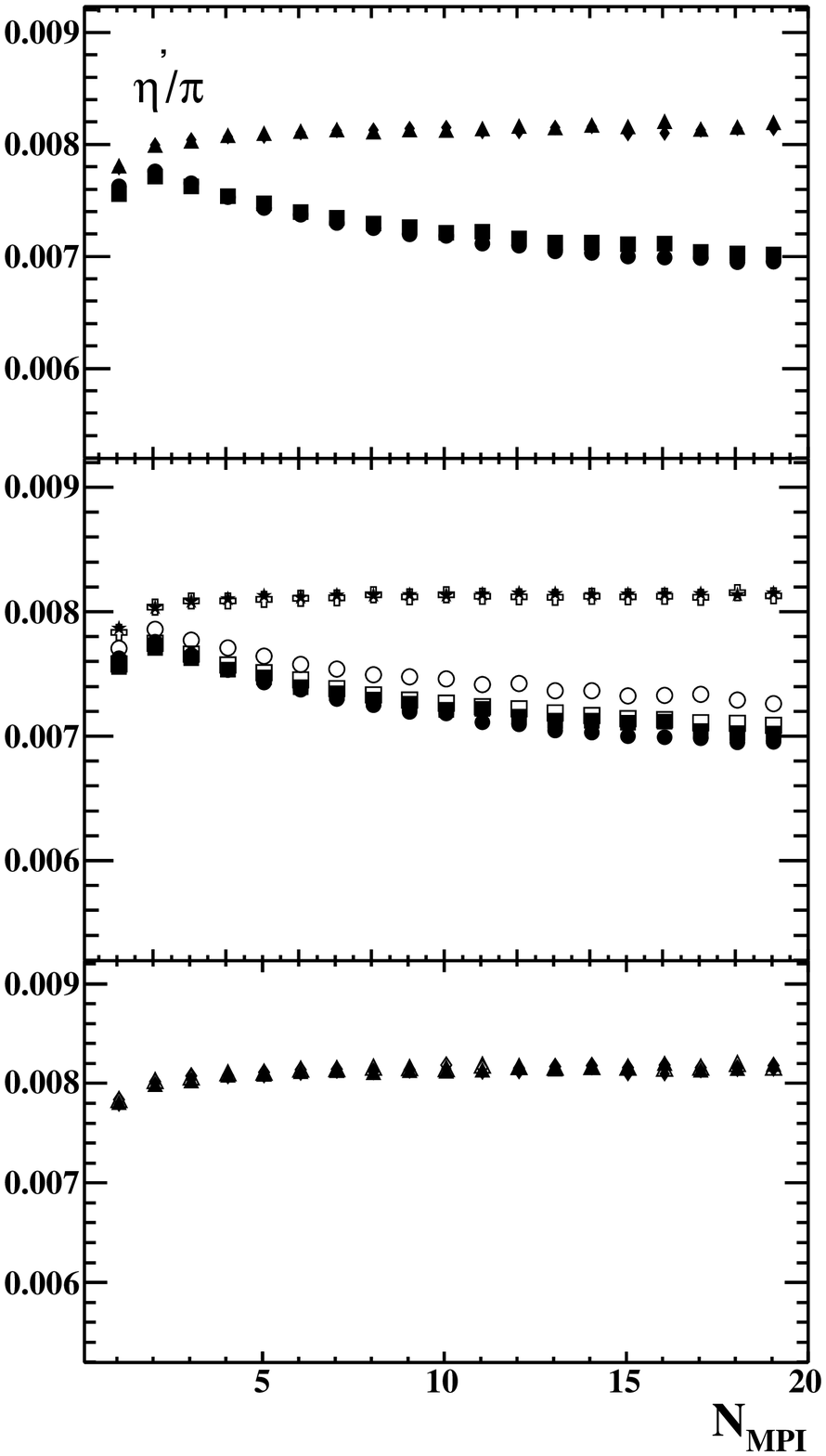}
\caption{}
\end{subfigure}
\begin{subfigure}[b]{0.32\textwidth}
\includegraphics[width=\textwidth]{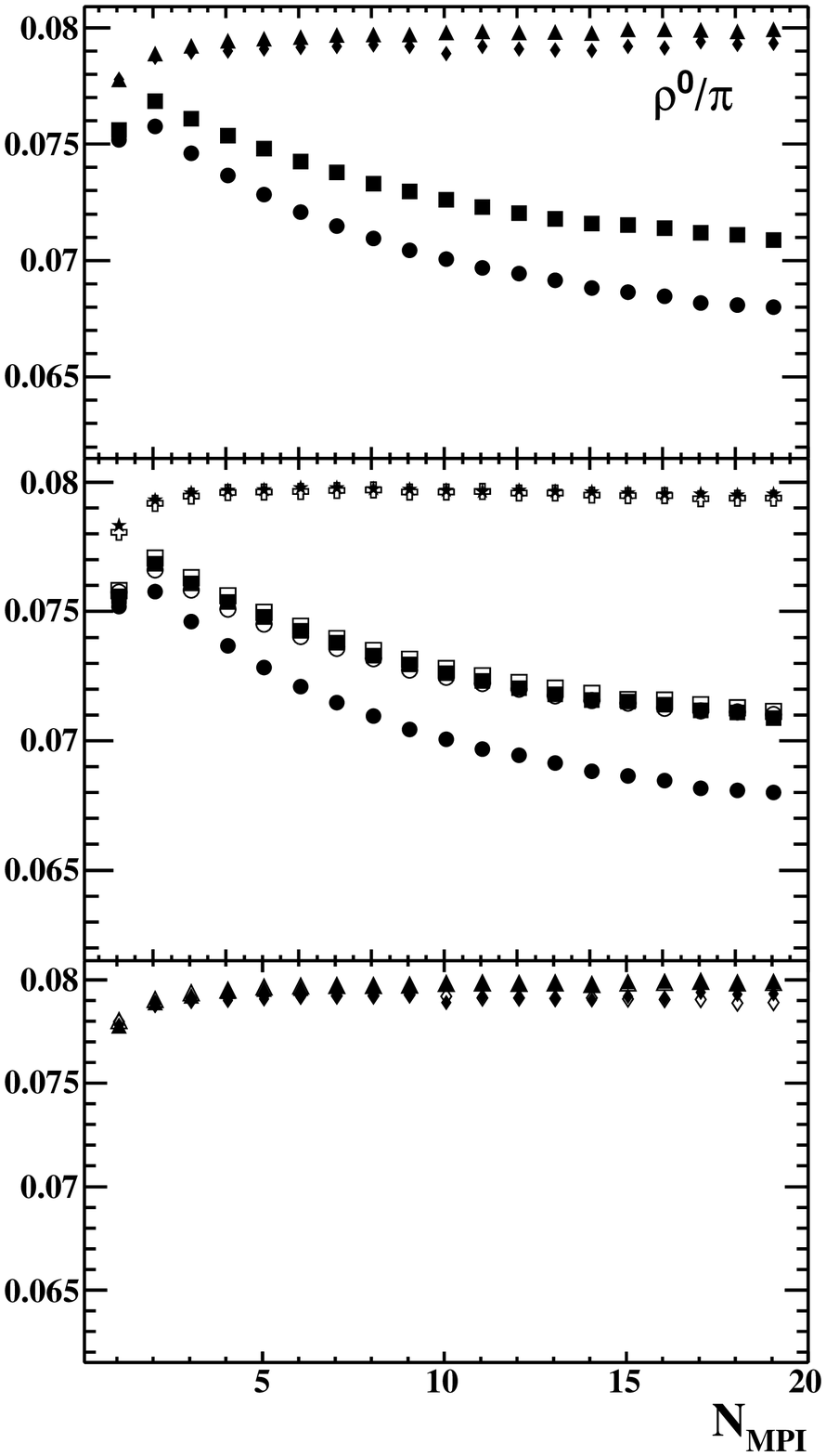}
\caption{}
\end{subfigure}
\caption{ $p_{T}$-integrated ratio of (a) $\mathrm{K^{*0}/K}$, (b) $\mathrm{\eta^{'}/\pi}$ and (c) $\mathrm{\rho^{0}/\pi}$ as a function of number of MPIs for p$-$p collisions. 
(Upper panel) The ratios are compared for two different modes of CR (CR-off, CR(0) and CR(1)) with (and without) RH in p$-$p collisions at $\sqrt{s}$ = 7 TeV. 
(Middle panel) The ratios are compared for  p$-$p collisions at $\sqrt{s}$ = 7 TeV and 13 TeV for CR-off, CR(0) and CR(1) with RH. (Bottom panel) Comparison of ratios for  p$-$p collisions at $\sqrt{s}$ = 7 TeV and 13 TeV for two other modes namely  CR off and CR(0) without RH.}
\label{comparemeson}
\end{figure*}

\begin{figure*}
\centering
\begin{subfigure}[b]{0.495\textwidth}
\includegraphics[width=\textwidth]{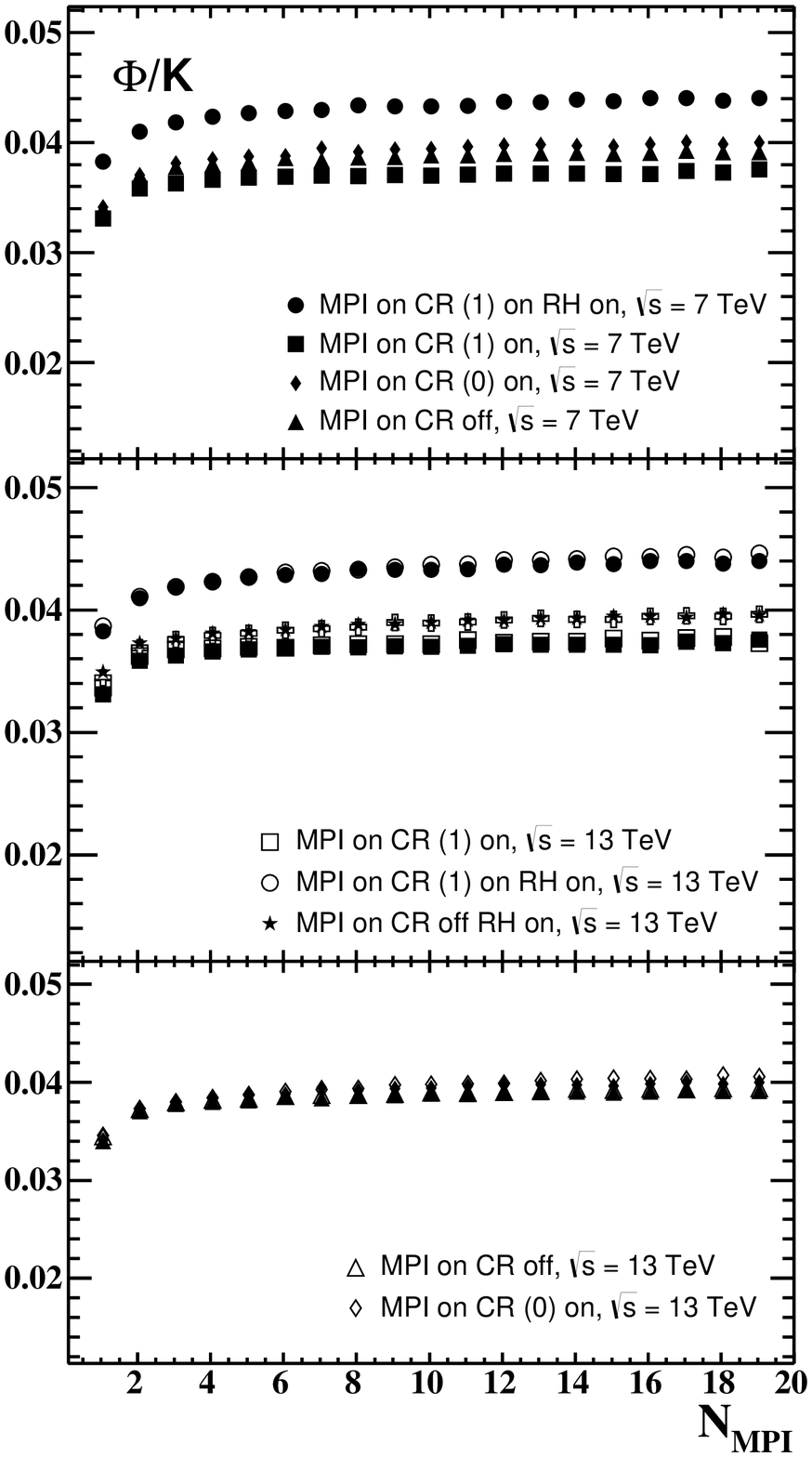}
\caption{}
\end{subfigure}
\hfill
\begin{subfigure}[b]{0.495\textwidth}
\includegraphics[width=\textwidth]{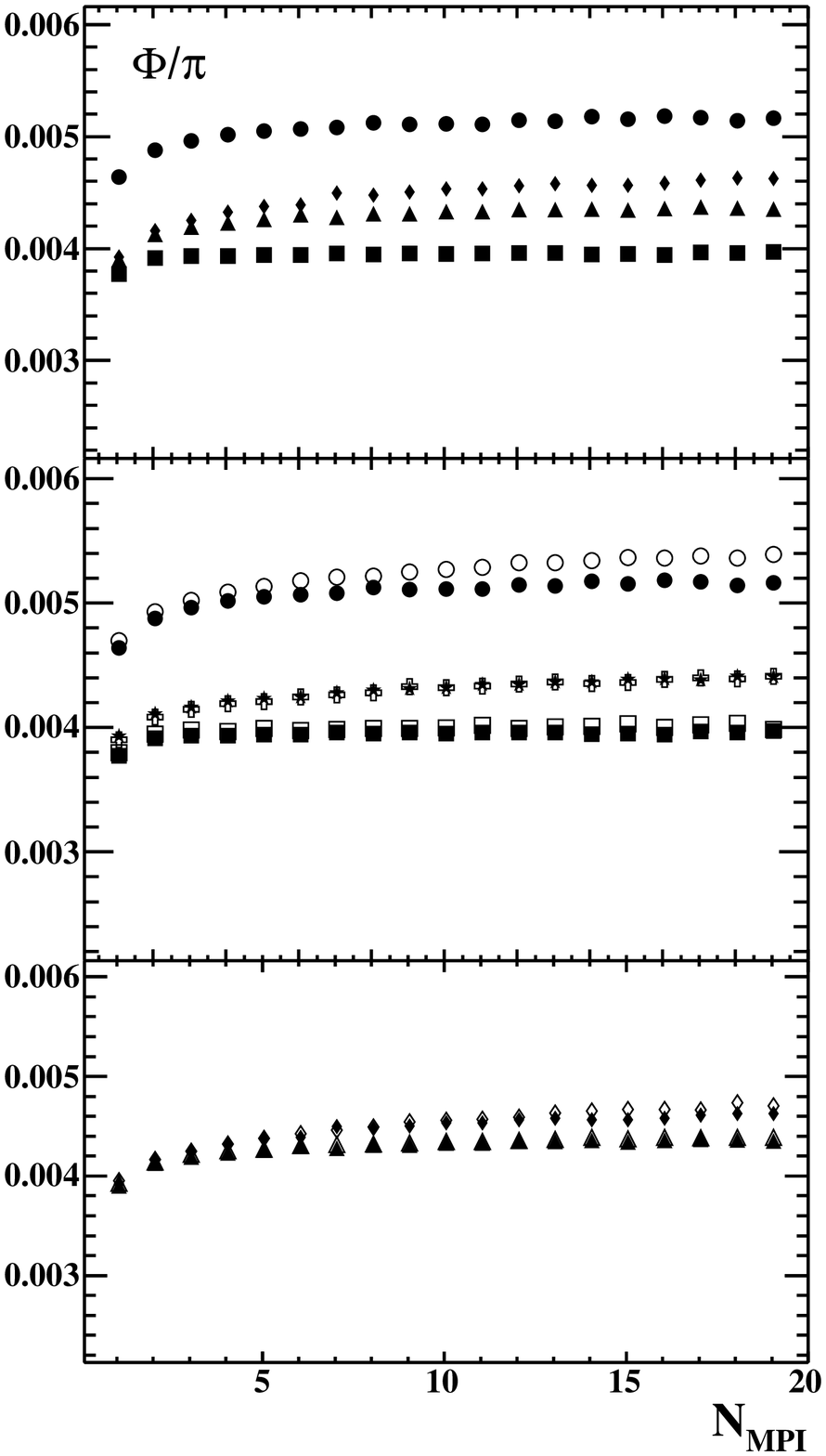}
\caption{}
\end{subfigure}
\caption{$p_{T}$-integrated ratio of (a) $\mathrm{\phi/K}$ and (b) $\mathrm{\phi/\pi}$ as a function of number of MPIs in p$-$p collisions. 
(Upper panel) The ratios are compared for two different modes of CR (CR-off, CR(0) and CR(1)) with (and without) RH in p$-$p collisions at $\sqrt{s}$ = 7 TeV. (Middle panel) The ratios are compared for p$-$p collisions at $\sqrt{s}$ = 7 TeV and 13 TeV for CR-off, CR(0) and CR(1) with RH.(Bottom panel) Comparison of ratios for  p$-$p collisions at $\sqrt{s}$ = 7 TeV and 13 TeV for two other modes namely  CR off and CR(0) without RH.}
\label{phiratio}
\end{figure*}

\section{MESONIC RESONANCES}

Figure \ref{comparemeson} shows the yield ratios of $\mathrm{K^{*0}/K}$, $\mathrm{\rho^{0}/\pi}$, $\mathrm{\eta^{'}/\pi}$ as a function of $N_{MPI}$ for p$-$p collisions at $\sqrt{s}$ = 7 TeV and 13 TeV. In the upper panel of Figure \ref{comparemeson}(a), the ratio of $\mathrm{K^{*}/K}$ shows a gradual suppression for high multiplicity events for QCD-based color reconnection (CR(1)) when compared to the scenario where there is no color reconnection or there is only MPI-based color reconnection(CR(0)). The values obtained after enabling rope hadronization is similar to that of CR(1) and its effect is not significant. The middle panel and the lower panel compare the ratio for p$-$p collisions at $\sqrt{s}$ = 7 TeV and 13 TeV. There is no effect of increasing the beam energy on the ratios indicating that the suppression is comparable for similar event activity. The suppression is similar for both CR-off and  CR(0) mode.The enabling of CR leads to production of large number of shorter strings and the effect is more pronounced for CR(1) scheme \cite{resoprd}. Therefore, the  production of massive $\mathrm{K^{*0}}$ is suppressed compared to kaons in high multiplicity events as shorter strings fragment to less massive particles. The formation of ropes in combination with the CR(1) gives similar suppression as RH leads to an enhanced production of strange hadrons and baryons. The effect of RH is not expected to be visible for $\mathrm{K^{*0}/K}$ as the quark content is same for both the particles.
We further investigated the yield ratio of $\mathrm{\eta^{'}/\pi}$ as a function of number of MPIs for p$-$p collisions at $\sqrt{s}$ =  7 TeV and 13 TeV in Figure \ref{comparemeson} (b). Both of the considered particles in the ratio are spin-0 particles with $\mathrm{\eta^{'}}$ being the heavier one compared to $\pi$. A similar trend like $\mathrm{K^{*0}/K}$ suppression is observed which can be attributed to the production of pions being favored more than $\eta^{'}$ due to formation and fragmentation of shorter strings.
The variation of $\mathrm{\rho^{0}/\pi}$ with multiplicity is shown in Figure \ref{comparemeson}(c). Surprisingly, a larger suppression in the yield ratio is observed for rope hadronization in combination with CR(1) for high multiplicity events. Since $\mathrm{\rho^{0}}$ is not a strange hadron, enabling RH further reduces its production in favor of other strange hadrons or baryons in high multiplicity events. However, a similar suppression was not seen for $\mathrm{K^{*0}/K}$ on enabling RH as 
both the particles are strange hadrons. 
The yield ratios of $\mathrm{\phi/K}$ and $\mathrm{\phi/\pi}$ have also been considered as a function of event activity as illustrated in Figure \ref{phiratio} (a) and (b), respectively. $\phi$ is a spin-1 resonance particle with relatively longer life time. 
Although $\phi$ consists of $s\bar{s}$ quarks, the yield ratios are enhanced due to the production of more strange (and anti-strange) quark pairs due to formation of color ropes. The enhancement is more pronounced for CR(1) in combination with rope hadronization, and also the enhancement in the yield ratios are independent of event activity and energies. It is interesting to note that the ratio for CR(1) is consistently smaller than CR(0) and CR-off. This can be due to the production of lower mass particles as CR(1) favors formation of shorter mass strings. 
\begin{figure}[h!]
\includegraphics[width=1.0\linewidth]{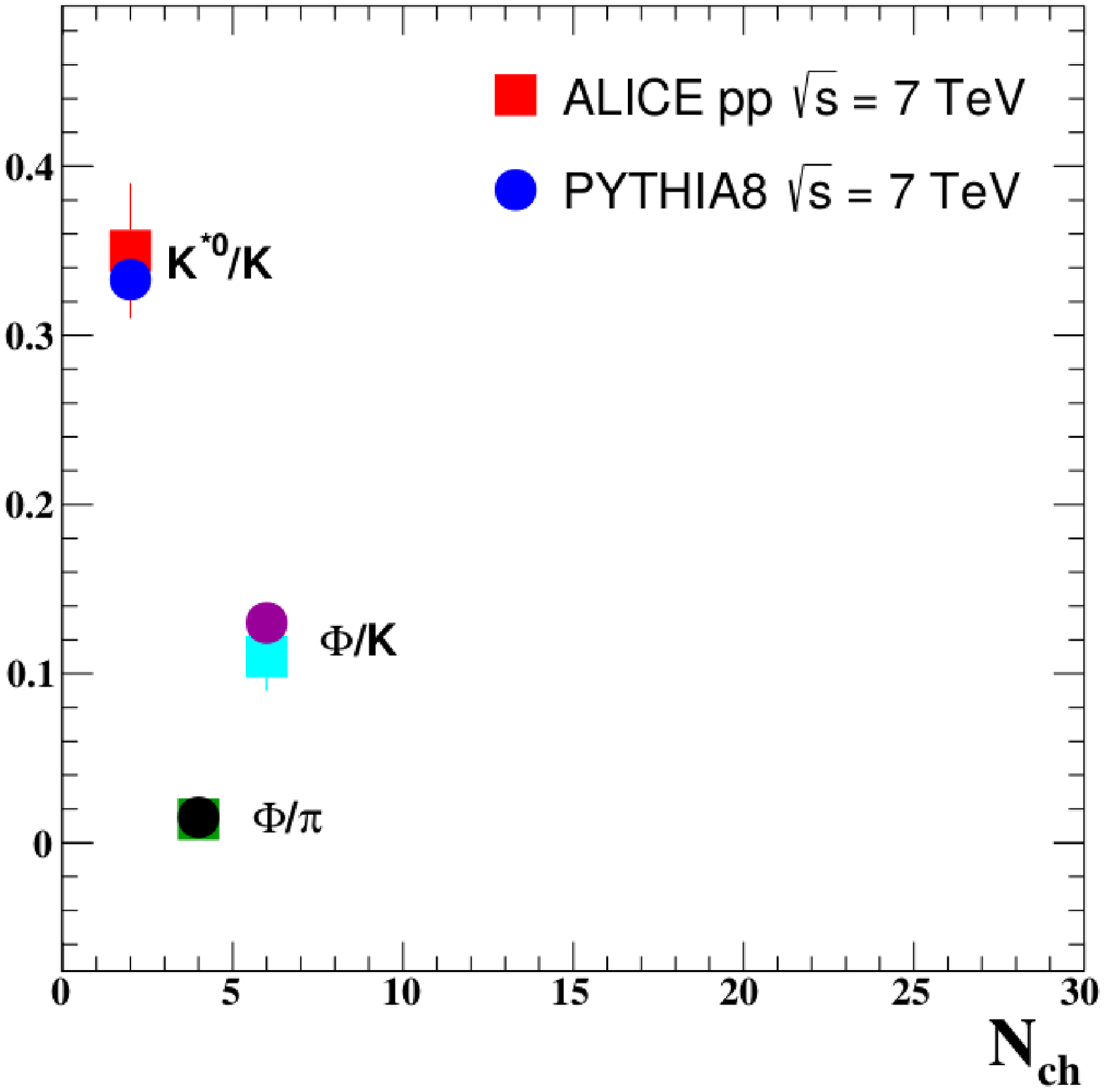}
\caption{The $p_{T}$-integrated ratio as a function of charged particle multiplicity in proton-proton collisions at 7 TeV \cite{alice}. }
\label{data}
\end{figure}

Figure \ref{data} shows the comparison of  $\mathrm{K^{*0}/K}$, $\mathrm{\phi/K}$, and $\mathrm{\phi/\pi}$ ratio obtained from Pythia 8 (with CR(1) and RH on) with the recent ALICE measurements in p$-$p collisions at $\sqrt{s}$ = 7 TeV. The solid circles show the yield ratios obtained from Pythia 8 while the solid squares represent the measured data. The values obtained from Pythia 8 are in good agreement with the measured data within the uncertainties which indicates that the microscopic processes like rope hadronization and color reconnection play an important role in describing the particle production mechanism at high energies.

\begin{figure*}
\centering
\begin{subfigure}[b]{0.49\textwidth}
\includegraphics[width=\textwidth]{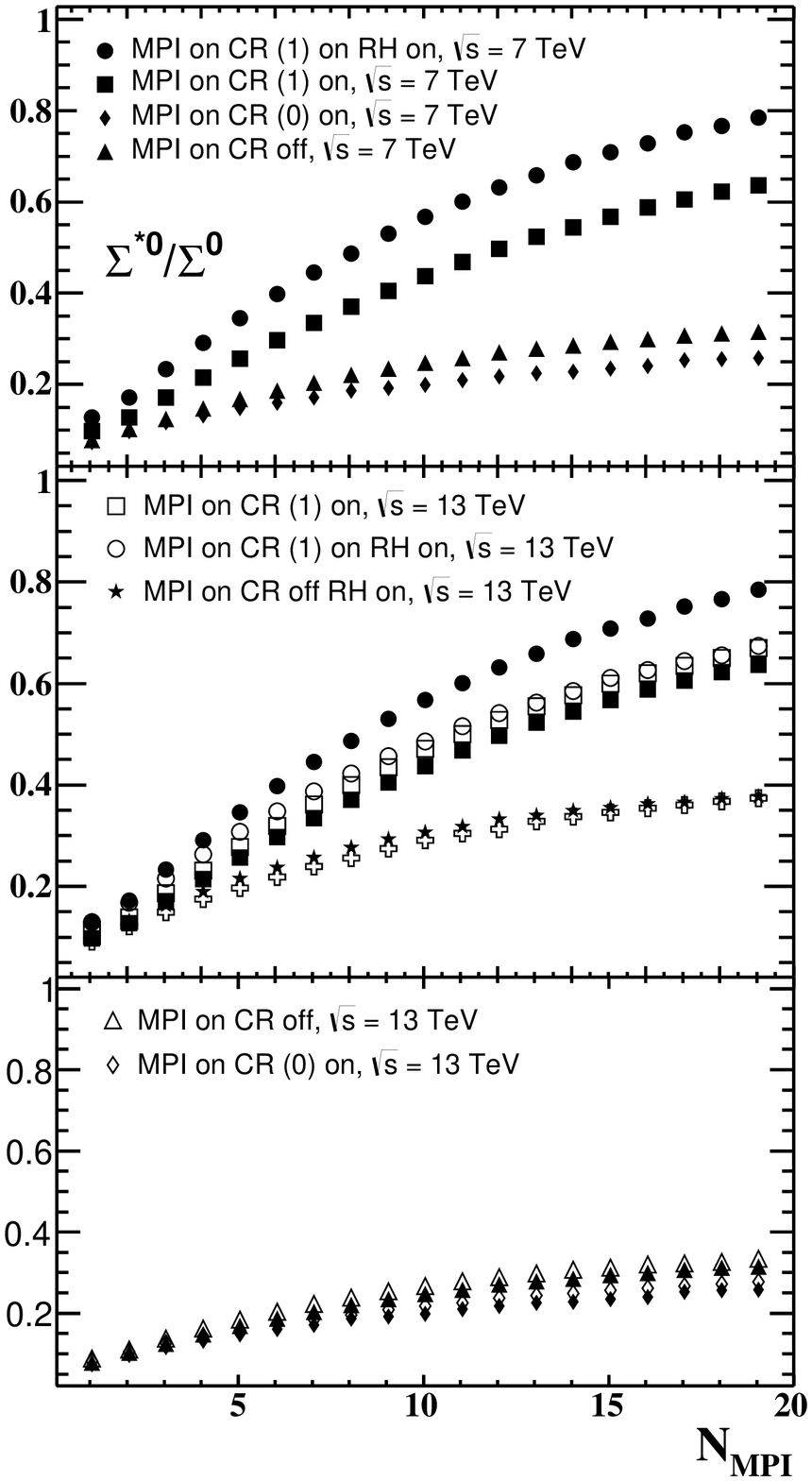}
\caption{}
\end{subfigure}
\hfill
\begin{subfigure}[b]{0.49\textwidth}
\includegraphics[width=\textwidth]{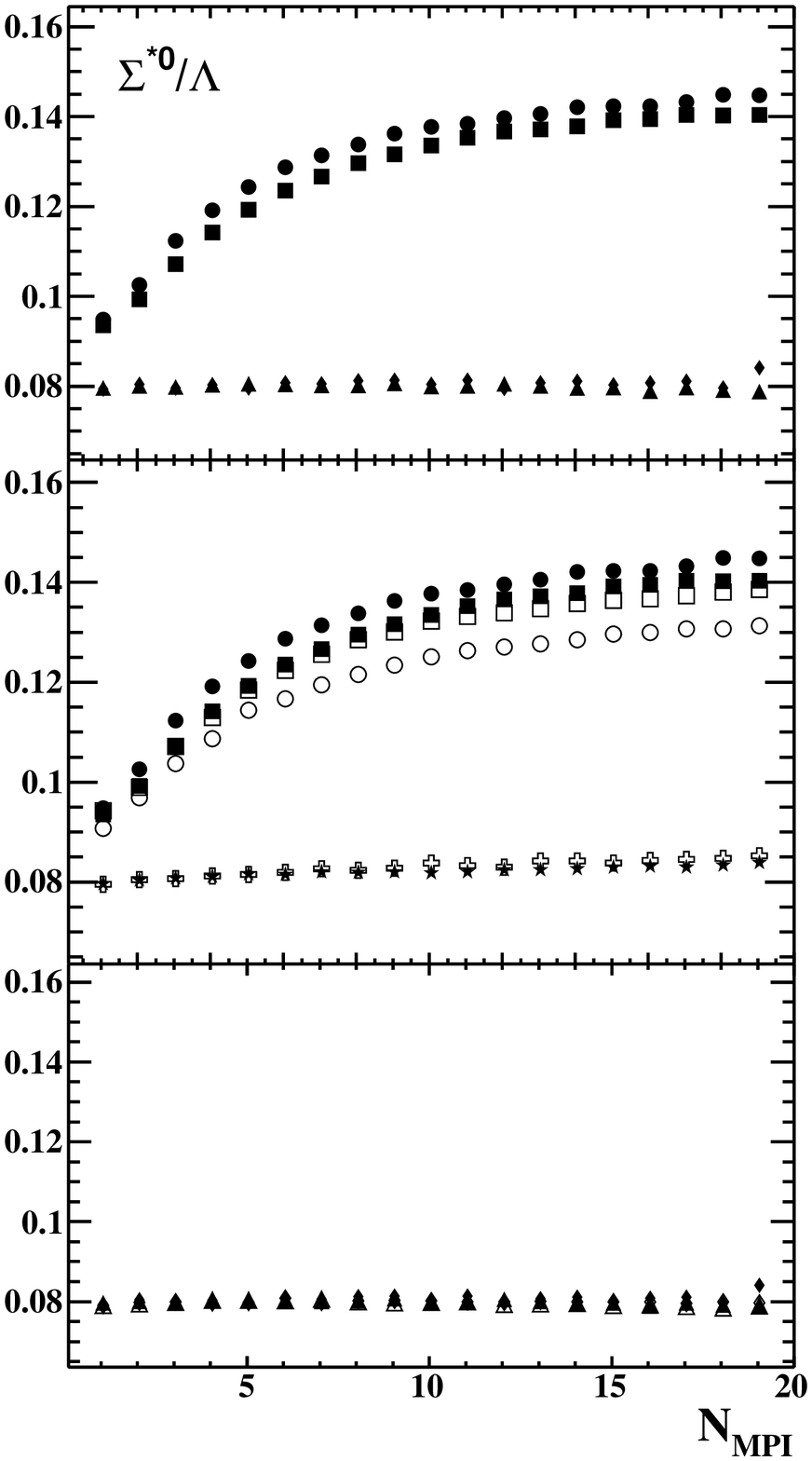}
\caption{}
\end{subfigure}
\caption{$p_{T}$-integrated ratio of (a) $\mathrm{\Sigma^{0*}/\Sigma}$ and (b) $\mathrm{\Sigma^{0*}/\Lambda}$ as a function of number of MPIs in p$-$p collisions.
(Upper panel) The ratios are compared for two different modes of CR (CR-off, CR(0) and CR(1)) with (and without) RH in p$-$p collisions at $\sqrt{s}$ = 7 TeV. (Middle panel) The ratios are compared for  p$-$p collisions at $\sqrt{s}$ = 7 TeV and 13 TeV for CR-off, CR(0) and CR(1) with RH.(Bottom panel) Comparison of ratios for  p$-$p collisions at $\sqrt{s}$ = 7 TeV and 13 TeV for two other modes namely  CR-off and CR(0) without RH.}
\label{delta}
\end{figure*}

\begin{figure*}
\centering
\begin{subfigure}[b]{0.49\textwidth}
\includegraphics[width=\textwidth]{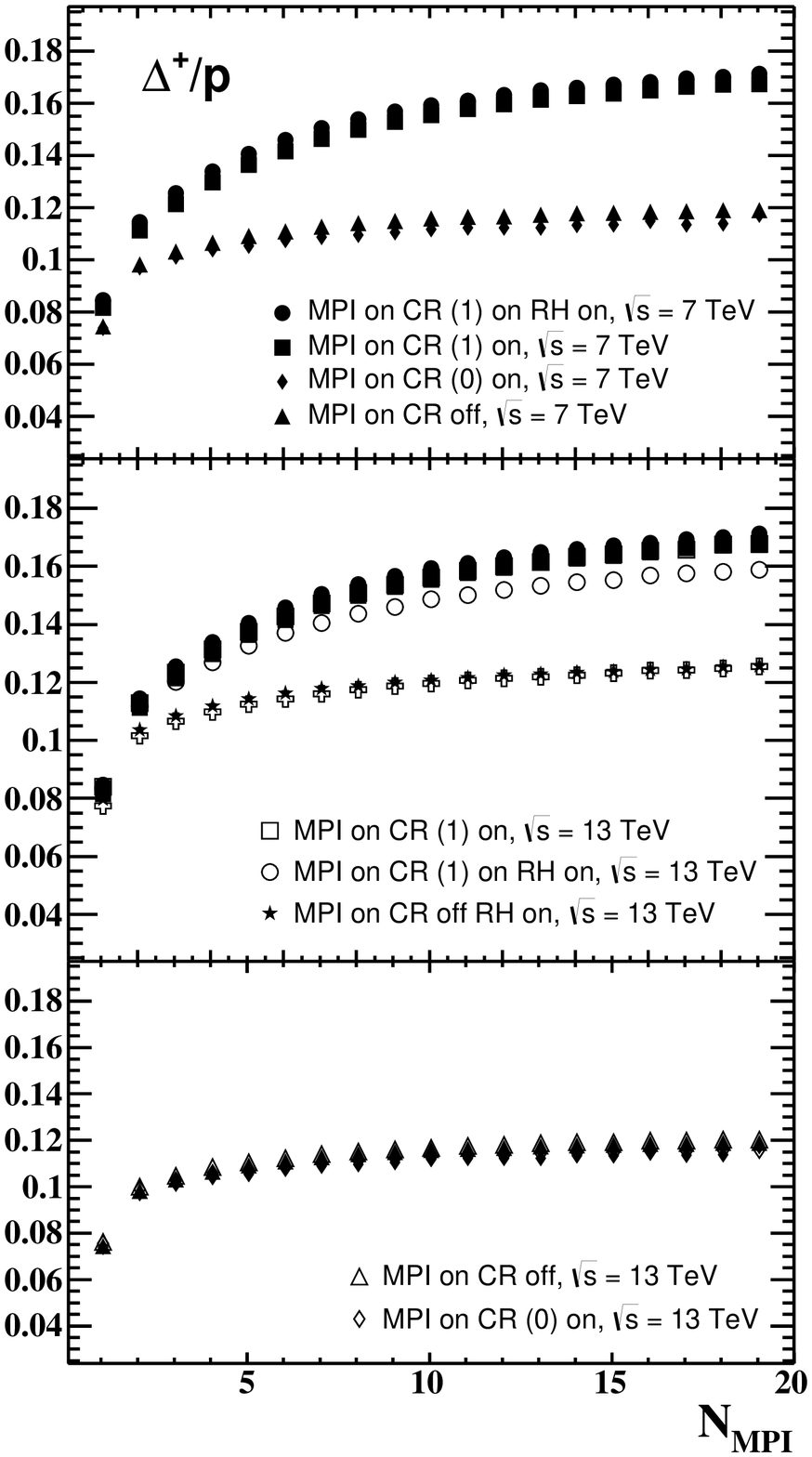}
\caption{}
\end{subfigure}
\begin{subfigure}[b]{0.49\textwidth}
\includegraphics[width=\textwidth]{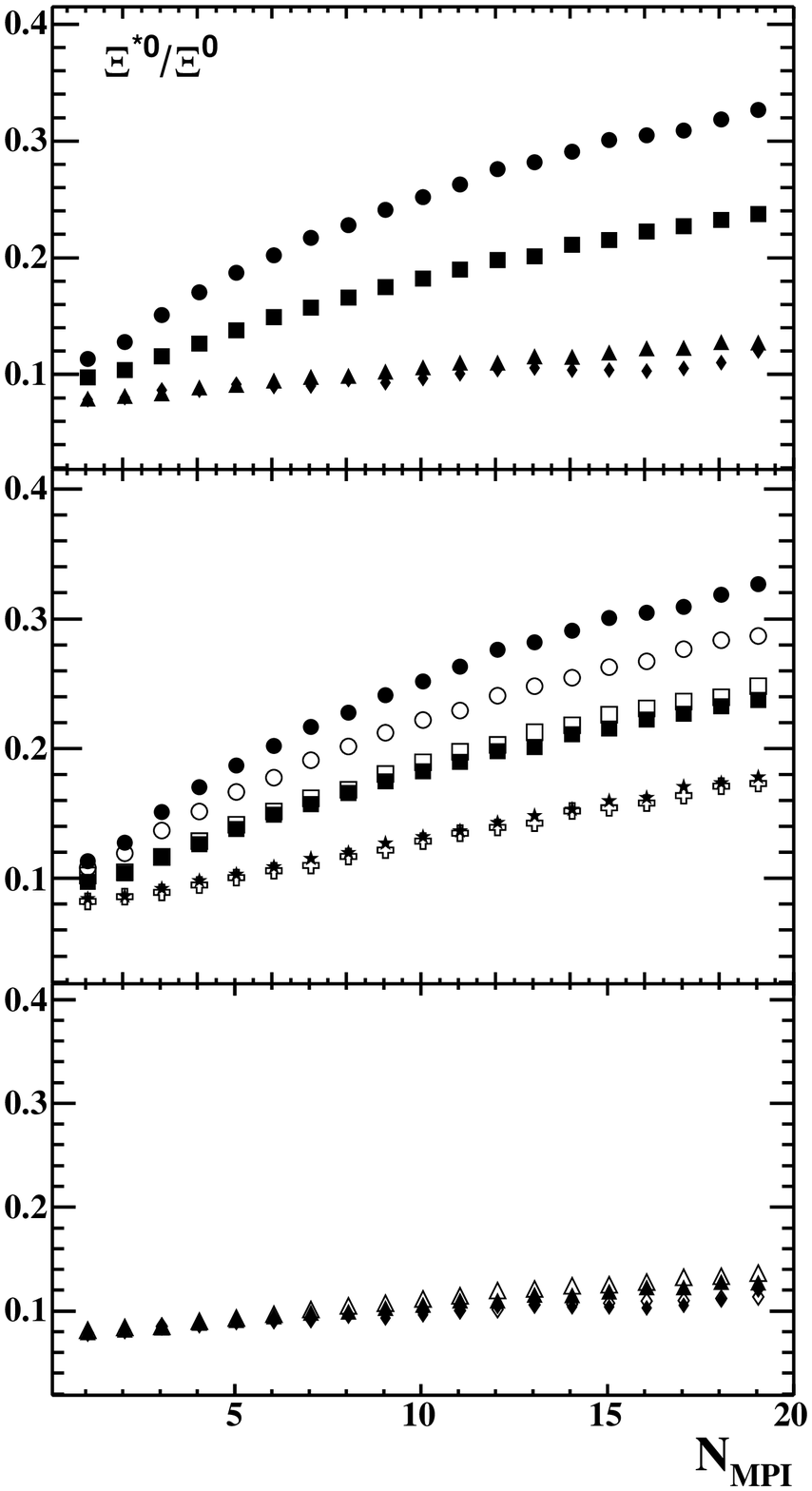}
\caption{}
\end{subfigure}
\caption{$p_{T}$-integrated ratio of (a) $\mathrm{\Delta^{+}/p}$ and (b) $\mathrm{\Xi^{0*}/\Xi^{0}}$ as a function of  number of MPIs in p$-$p collisions. (Upper panel) The ratios are compared for two different modes of CR (CR-off, CR(0) and CR(1)) with (and without) RH in p$-$p collisions at $\sqrt{s}$ = 7 TeV. (Middle panel) The ratios are compared for  p$-$p collisions at $\sqrt{s}$ = 7 TeV and 13 TeV for CR-off, CR(0) and CR(1) with RH.(Bottom panel) Comparison of ratios for  p$-$p collisions at $\sqrt{s}$ = 7 TeV and 13 TeV for two other modes namely  CR off and CR(0) without RH.}
\label{sigma}
\end{figure*}

\section{Baryonic Resonances }
The particle ratios involving the baryonic resonances have also been investigated by studying  the $\mathrm{\Delta^{+}/p}$, $\mathrm{\Sigma^{0*}/\Lambda}$, $\mathrm{\Sigma^{0*}/\Sigma^{0}}$, and $\mathrm{\Xi^{0*}/\Xi^{0}}$. Figure \ref{sigma} (a) and (b) show the variation of yield ratios of $\mathrm{\Sigma^{0*}/\Sigma^{0}}$ and $\mathrm{\Sigma^{0*}/\Lambda}$ with respect to number of MPIs. It is interesting to observe the evolution of $\mathrm{\Sigma^{0*}/\Sigma^{0}}$ and  $\mathrm{\Sigma^{0*}/\Lambda}$ with event activity as the particles in the denominator have identical quark content but differ in spin. $\Sigma^{0}$ belongs to the isospin triplet while $\Lambda^{0}$ is a singlet. We observe an enhancement in the ratio of $\mathrm{\Sigma^{0*}/\Sigma^{0}}$ and $\mathrm{\Sigma^{0*}/\Lambda}$ with CR(1) but the enhancement is more pronounced in case of rope hadronization for the former while no visible effect is seen for the later. This shows that production of baryons belonging to higher spin multiplet is favored when rope hadronization is switched on. This leads to more production of $\Sigma^{0*}$ compared to $\Sigma^{0}$. The effect is not seen for $\mathrm{\Sigma^{*0}/\Lambda}$ as they have differ in spin and isospin. The effect has no significant dependence on energies as can be clearly seen in the middle and lower panels of Fig.\ref{sigma} (a) and (b). 
Figure \ref{delta} (a) shows the $\Delta^{+}/p$ ratio as a function of event activity. $\Delta^{+}$ is an interesting baryon as it has an extremely short lifetime, it is a non-strange baryon with higher spin ($3/2$) and has same quark content as proton. They differ in mass and spin. The ratio is observed to increase with event activity when CR(1) is enabled and rope hadronization has no observable effect on the ratios. This observed enhancement is in agreement with the effects of CR(1) as CR(1) produces more baryons of higher spin. The enabling of rope hadronization has no effect on enhancement as both the particles are non-strange particles. 
Furthermore, $\mathrm{\Xi^{0*}/\Xi^{0}}$ ratio as shown in Figure \ref{delta} (b) was also investigated as $\Xi$ is a doubly 
strange baryon. Both the particles have  same quark content but differ in mass and spin. The effect of CR(1) leads to an enhancement because of favorable production of higher spin baryons. Additionally, the color rope formation increases the production of strange quarks and favors the strange baryon production with higher spin which is responsible for further enhancement in the ratio with RH. The collisional energy independence is also observed for both the ratios as illustrated in middle and lower panels of Figure \ref{delta} (a) and (b).

For all the resonances, both mesonic and baryonic, the effect of the rope hadronization was also studied without the mechanism of color reconnections and it has been observed that it has no effect on the ratios which asserts that color reconnection plays the most important role in the observed suppression (or enhancement) of resonance to non-resonance ratios. The obtained results  serve as a baseline for future measurements and provide an alternate description for suppression (or enhancement) of resonance to non-resonance ratios  based on microscopic processes active in the partonic domain.

\section{Summary}
The production of resonance particles in p$-$p collisions has been studied at  $\sqrt{s}$ = 7 TeV and 13 TeV using Pythia 8 event generator. 
The ratio  of yield of resonance to non-resonance particle with similar quark content has been obtained within the framework of microscopic processes like color reconnections and rope hadronization.  The ratios for various species of mesonic and baryonic resonances are studied as a function of number of MPIs to study the effect of the considered microscopic processes. A suppression is observed for mesonic resonance ratios like $\mathrm{K^{*0}/K}$, $\mathrm{\eta^{'}/\pi}$ and $\mathrm{\rho^{0}/\pi}$ for high event activity when the QCD-based color reconnection is enabled. The observation can be attributed to the formation of shorter strings which fragment into lower mass particles. The yield ratios for $\mathrm{\phi/K}$ and $\mathrm{\phi/\pi}$ increases for QCD-based CR and RH compared to other modes. The production of $\phi$ increases with the production of more strange (and anti-strange) quarks with color ropes. The model predicts an enhancement in the yield ratios for baryonic resonances. The ratios were found to be independent of the collision energy of the system. These 
predictions can serve as a baseline study for future measurements and can be an alternative explanation of the observed resonance suppression and regeneration based on processes present in partonic phase.   

\begin{acknowledgements}
The authors would like to thank Department of Science and Technology (DST), India for supporting the present work.
\end{acknowledgements}

\end{document}